\documentclass[twocolumn,showpacs,preprintnumbers,amsmath,amssymb,pra,aps,superscriptaddress]{revtex4-1}

\usepackage{graphicx}
\usepackage{multirow}
\usepackage{amsmath}
\usepackage{soul}
\usepackage{verbatim}
\usepackage{bm}
\usepackage{outlines}
\usepackage{siunitx}
\usepackage[space]{grffile}
\usepackage[colorlinks=true, linkcolor=blue, citecolor=gray]{hyperref}
\usepackage[capitalize]{cleveref}
\usepackage{xr}
\usepackage{color}
\usepackage{tikz}
\usetikzlibrary{calc, positioning}

\newcommand{\<}{\langle}
\renewcommand{\>}{\rangle}

\begin{document}

\title{Majorana-based fermionic quantum computation}
\author{T. E. O'Brien}
\affiliation{Instituut-Lorentz, Universiteit Leiden, P.O. Box 9506, 2300 RA Leiden, The Netherlands}
\author{P. Ro\.{z}ek}
\affiliation{QuTech, Delft University of Technology, P.O. Box 5046, 2600 GA Delft, The Netherlands}
\affiliation{Kavli Institute of Nanoscience, Delft University of Technology, P.O. Box 5046, 2600 GA Delft, The Netherlands}
\author{A. R. Akhmerov}
\affiliation{Kavli Institute of Nanoscience, Delft University of Technology, P.O. Box 5046, 2600 GA Delft, The Netherlands}
\date{\today}

\begin{abstract}
Because Majorana zero modes store quantum information non-locally, they are protected from noise, and have been proposed as a building block for a quantum computer.
We show how to use the same protection from noise to implement universal fermionic quantum computation.
Our architecture requires only two Majoranas to encode a fermionic quantum degree of freedom, compared to alternative implementations which require a minimum of four Majoranas for a spin quantum degree of freedom.
The fermionic degrees of freedom support both unitary coupled cluster variational quantum eigensolver and quantum phase estimation algorithms, proposed for quantum chemistry simulations.
Because we avoid the Jordan-Wigner transformation, our scheme has a lower overhead for implementing both of these algorithms, and the simulation of Trotterized Hubbard Hamiltonian in $\mathcal{O}(1)$ time per unitary step.
We finally demonstrate magic state distillation in our fermionic architecture, giving a universal set of topologically protected fermionic quantum gates.
\end{abstract}

\maketitle

Particle exchange statistics is a fundamental quantum property that distinguishes commuting spin or qubit degrees of freedom from anticommuting fermions, despite single particles in both systems only having two quantum states.
Different exchange statistics cause a different set of Hamiltonian terms to be local, or even physically possible.
For example, although it is Hermitian, the linear superposition of a fermionic creation and annihilation operator $c+c^\dagger$ never occurs as a Hamiltonian term in nature due to violating fermion parity conservation, whilst spin systems have no such restrictions.
Despite these differences, it is possible to simulate fermions using qubits and vice versa~\cite{bravyi_fermionic_2002}.
Such simulation necessarily incurs overhead because of the need to transform local fermion operators into non-local qubit ones by using, for example, the Jordan-Wigner transformation.
Because quantum simulation of the electronic structure of molecules is a promising application of quantum computation~\cite{reiher_elucidating_2017}, much recent work focused on minimizing this overhead of simulating fermionic Hamiltonians with qubits~\cite{Has14,kivlichan_quantum_2017,Bab17}.

Majorana zero modes (also Majorana modes or just Majoranas) are non-abelian particles, with two Majoranas combining to form a single fermion (see e.g. Refs.~\cite{alicea_new_2012,beenakker_search_2013,leijnse_introduction_2012} for a review).
Spatially separating two Majoranas protects this fermionic degree of freedom, and provides a natural implementation of a topological quantum computer~\cite{kitaev_unpaired_2001-1,sarma_majorana_2015-1}.
Further, conservation of fermion parity prevents creating a superposition between the two different parity states of two Majoranas, and therefore most of the existing proposals combine 4 Majoranas with a fixed fermion parity into a single qubit.

Fermionic quantum computation~\cite{bravyi_fermionic_2002} was so far not actively pursued because of the lack of known ways to protect fermionic degrees of freedom from dephasing.
We observe that Majoranas naturally offer this protection, while in addition providing a platform for implementing quantum chemistry algorithms.
We therefore show that for the problem of simulating fermionic systems on a Majorana quantum computing architecture, it is both possible and preferable to use fermions composed from pairs of Majoranas instead of further combining pairs of these fermions to form single qubits.
Formulating fermionic quantum simulation algorithms in terms of fermions imposes the fermion parity conservation at the hardware level, and prohibits a large class of errors bringing the simulator out of the physical subspace.
Furthermore, working natively with fermions, we remove the need for the Jordan-Wigner (or related) transformation to map a fermionic problem to a spin system.
When simulating a typical quantum chemistry Hamiltonian, our approach results in a more dense encoding of the computational degrees of freedom.
The benefit from using the fermionic degrees of freedom becomes more important in simulating local fermionic Hamiltonians, such as the Hubbard model, allowing the simulation of unitary time evolution in $\mathcal{O}(1)$ time per Trotter step, and further reducing the cost of pre-error-correction quantum simulation~\cite{dal16}.
Finally, we show how to apply the known magic state distillation protocol in fermionic quantum computation.
Combined with the recent realization of the fermionic error correction~\cite{li_fault-tolerant_2017} this provides a fault-tolerant fermionic quantum computer.

\begin{figure}
\includegraphics[width=0.5\textwidth]{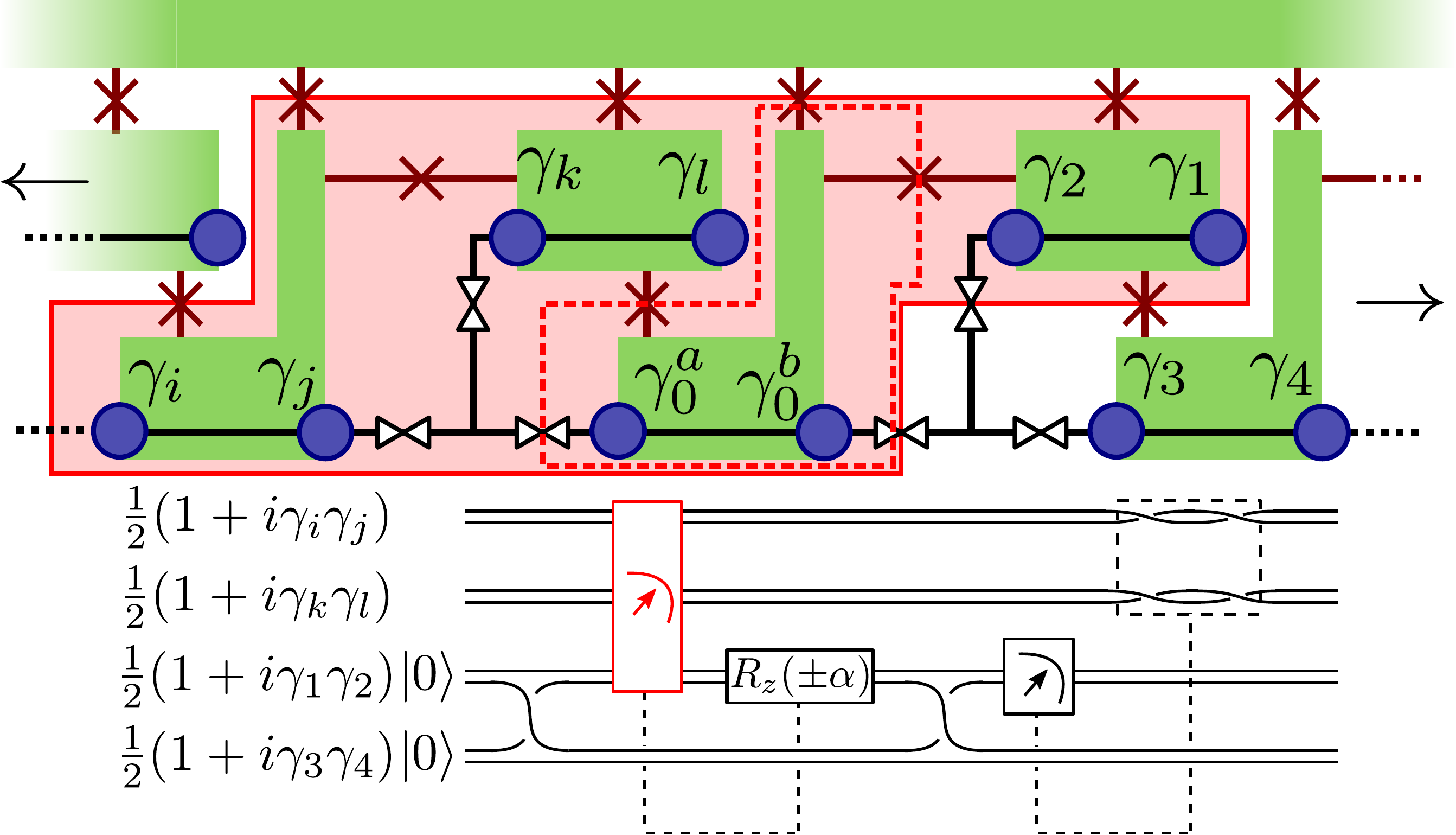}
\caption{\label{fig:4fold_rotation}Top: a $1D$ implementation of a Majorana circuit. Majoranas (blue dots) occur at either the edge of a nanowire (black line) or as it crosses the boundary of a superconductor (light green).
Josephson junctions (red crossed lines) connect superconducting islands to a common base, allowing for parallel joint parity measurements.
Fully-tunable T-junctions (valve symbols) allow for a computational Majorana to be shifted from one end of any coupled set of itself and two braiding ancillas (prepared in a known state) to another end. Bottom: an implementation of a weight-four Majorana rotation (Eq.~\ref{eq:Majorana_rotation}) using the labeled qubits in the design. The operation of individual circuit elements is listen in Table.~\ref{tab:legend}. The highlighted parity measurement is performed by isolating the highlighted area of the architecture via tunable Josephson junctions, and measuring the total charge parity. This requires a separate preparation of the Majoranas $\gamma_0^a$ and $\gamma_0^b$ (dashed red box) in the $i\gamma_0^a\gamma_0^b=1$ state (which is also required to use these as spare sites for braiding).}
\end{figure}

Our approach relies on the known set of ingredients to perform universal operations with Majorana states~\cite{hyart_flux-controlled_2013-1}: controllable Josephson junctions, direct Majorana coupling, and Coulomb energy.
A possible architecture implementing a Majorana-based fermionic quantum processor is shown in Fig.~\ref{fig:4fold_rotation}.
Because our system cannot be separated into blocks with a fixed fermion parity, the protection of the quantum degrees of freedom is only possible if different parts of the system are connected to a common superconducting ground~\footnote{The need to use a common superconducting ground makes it impossible to utilize the partial protection from quasiparticle poisoning by applying Coulomb blockade to superconducting islands containing individual qubits~\cite{karzig_scalable_2017}.}.
Turning off some of the Josephson junctions (these may be either flux-controlled SQUIDs or gate-controlled \cite{larsen_semiconductor-nanowire-based_2015,de_lange_realization_2015}) then isolates a part of the system, and generates a Coulomb interaction \cite{fu_electron_2010,van_heck_coulomb_2011-1}
\begin{equation}
  \label{eq:charging}
  H_C = i^{N/2} E_C \prod_k^N \gamma_k,
\end{equation}
that couples all the Majorana modes $\gamma_i$ belonging to the isolated part of the system with the charging energy $E_C$.
An example of such coupling acting on $8$ Majorana modes is shown by a red box in Fig.~\ref{fig:4fold_rotation}.
Finally, gate-controlled T-junctions exert the interaction
\begin{equation}
  \label{eq:majorana coupling}
  H_M = i E_M \gamma_j \gamma_k,
\end{equation}
on any two Majorana modes coupled by a T-junction, with $E_M$ the Majorana coupling energy.

Controllable pairwise interactions between Majorana modes~\cite{sau_controlling_2011,heck_coulomb-assisted_2012} or two-Majorana parity measurements~\cite{knapp_nature_2016} allow the implementation of braiding, while the joint readout of the fermionic parity of more than 2 Majorana modes generates the rest of the Clifford group~\cite{hyart_flux-controlled_2013-1}.
Finally, a diabatic pulse of a two-Majorana coupling implements an unprotected phase gate $e^{\theta\gamma_i\gamma_j}$.
We summarize these elementary gates that serve as a basis of our protocol in Table~\ref{tab:legend}.
This gate set is computationally universal within a fixed fermion parity sector~\cite{bravyi_fermionic_2002}.

\begin{table}
  \begin{ruledtabular}
    \begin{tabular}{m{2.5cm}|m{3cm}|m{2.8cm}}
      Name & Element & Operation\\ \hline
      Preparation & \vspace{.1cm} \centering{\includegraphics[width=2cm]{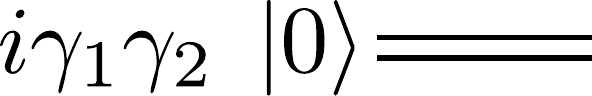}} & Prepare $\left(\begin{array}{c}1\\0\end{array}\right)$ \\ \hline
      Braiding & \vspace{.1cm} \centering{\includegraphics[width=2cm]{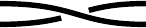}} & $\left(\begin{array}{cc}e^{i\pi/4} & 0 \\ 0 & e^{-i\pi/4}\end{array}\right)$\\ \hline
      Braiding &\vspace{.1cm}  \centering{\includegraphics[width=2cm]{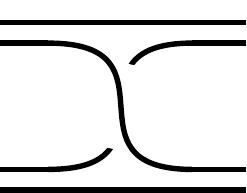}} & $\left(\begin{array}{cccc} 1 & 0 & 0 & -i\\ 0 & 1 & -i & 0 \\ 0 & -i & 1 & 0 \\ -i & 0 & 0 & 1\end{array}\right)$\\ \hline
      Rotation & \vspace{.1cm} \centering{\includegraphics[width=2cm]{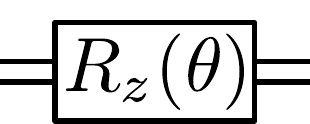}} & $\left(\begin{array}{cc}e^{i\phi} & 0 \\ 0 & e^{-i\phi}\end{array}\right)$ \\ \hline
      Measurement &\vspace{.1cm}  \centering{\includegraphics[width=2cm]{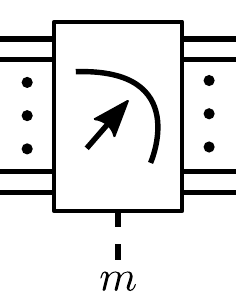}} & $\sum_{P (\phi)=m}|\phi \> \<\phi|$\\
    \end{tabular}
  \end{ruledtabular}
\caption{\label{tab:legend} Basic circuit elements we allow in our computation scheme. The above is sufficient to generate universal quantum computation in the single-parity sector. Computational degrees of freedom are formed by two Majoranas, and are therefore represented as a double line. Preparation, braiding, and measurement gates are assumed to be topologically protected. The $R_z (\theta)$ rotation is not topologically protected, but may be distilled via our magic state distillation protocol. The measurement projects our system onto a state of definite parity $P(\phi)$, being the sum $\sum_{i,j}\frac{1}{2}(1+i\gamma_i\gamma_j)$ of the pairs of Majoranas $\gamma_i,\gamma_j$ on islands connected to ground via Josephson junctions.}
\end{table}

The above gate set is sufficient to construct circuits for time evolution, quantum phase estimation (QPE), and a variational quantum eigensolver---the unitary coupled cluster ansatz (UCC).
Most fermionic systems have Hamiltonians constructed from twofold and fourfold fermionic terms:
\begin{equation}
H=\sum_{i,j}h_{i,j}\hat{f}^{\dag}_i\hat{f}_j+\sum_{i,j,k,l}\hat{f}^{\dag}_i\hat{f}^{\dag}_j\hat{f}_k\hat{f}_l.
\end{equation}
Here, $\hat{f}^{\dag}_i$ ($\hat{f}_i$) is the creation (annihilation) operator for an electron.
This is equivalent to a sum over 2 and 4-fold Majorana terms:
\begin{equation}
H=\sum_{i,j}ig_{i,j}\gamma_i\gamma_j+\sum_{i,j,k,l}g_{i,j,k,l}\gamma_i\gamma_j\gamma_k\gamma_l.
\end{equation}
Time evolution is performed by applying the Trotter expansion of the evolution operator $e^{iHt}$~\footnote{We have not discussed post-Trotter methods such as~\cite{ber11,ber17,pou17} in this work. However, these methods still require the Jordan-Wigner transformation or equivalent to represent a fermionic Hamiltonian on a qubit architecture. As such, they gain a similar advantage to the studied Trotterized evolution of $e^{iHt}$ from a Majorana-based fermion implementation.}:
\begin{equation}
e^{iH \Delta t} \underset{\Delta t \rightarrow 0}{\approx} \prod_{i,j}e^{-g_{i,j}\gamma_i\gamma_j \Delta t}\prod_{i,j,k,l}e^{ig_{i,j,k,l}\gamma_i\gamma_j\gamma_k\gamma_l \Delta t},
\end{equation}
and thus requires consecutive application of the unitary operators $e^{\theta\gamma_i\gamma_j}$ and $e^{i\theta\gamma_i\gamma_j\gamma_k\gamma_l}$.
We therefore introduce the weight-$2N$ Majorana rotation operator
\begin{equation}
\exp\left\{i\theta\prod_{n=1}^Ni\gamma_{2n-1}\gamma_{2n}\right\}\label{eq:Majorana_rotation},
\end{equation}
that forms the basis of all the algorithms we consider.

A Majorana rotation may be performed using a generic circuit with an additional four-Majorana ancilla qubit.
To demonstrate, the circuit of Fig.~\ref{fig:4fold_rotation} applies a Majorana rotation $e^{i\theta\gamma_i\gamma_j\gamma_k\gamma_l}$.
The same scheme implements weight-two Majorana rotations by removing Majoranas $\gamma_k$ and $\gamma_l$, and higher weight-$2N$ Majorana rotations by adding $2N-4$ more Majoranas to the correlated parity check and conditional final braiding.
The ancillary Majoranas $\gamma_0^a$ and $\gamma_0^b$ used for the braiding begin in the parity eigenstate $i\gamma_0^a\gamma_0^b=1$.
The eight-Majorana charge parity measurement $\gamma_i\gamma_j\gamma_k\gamma_l\gamma_0^a\gamma_0^b\gamma_2\gamma_1$ (implemented by isolating the circled superconducting islands in Fig.~\ref{fig:4fold_rotation}) therefore reduces to the 6-Majorana measurement highlighted in the circuit.
The unprotected rotation by the angle $\alpha=\theta+\frac{\pi}{2}$ both corrects an unwanted phase from the braiding of $\gamma_2$ and $\gamma_3$, and applies the non-Clifford rotation by $\theta$.

Quantum phase estimation requires the unitary evolution of a state (prepared across a set of system qubits) conditional on a set of ancilla qubits, which then have the eigenphases of the unitary operator encoded upon them~\cite{kit96}.
For the purposes of simulating quantum chemistry, a common choice of this operator is the time evolution operator, approximated by the Trotter expansion.
In App.~\ref{app:QPE_ancilla}, we show how to encode the ancilla qubit non-locally across an array of fermions, each of those controlling the unitary evolution of a local Hamiltonian term.
This reduces the requirements for QPE to consecutive operations of weight-four and weight-six Majorana rotations, with two Majoranas in each rotation belonging to an ancilla fermion.
In App.~\ref{app:TrotterN3} we show how this circuit is used to execute a single Trotter step for a fully-connected fourth-order Hamiltonian in $O(N^3)$ time.

Variational quantum eigensolvers prepare a trial state $|\psi(\vec{\theta})\rangle$ from a circuit depending on a set of variational parameters $\vec{\theta}$, which are then tuned to minimize the energy $\langle\psi(\vec{\theta})|H|\psi(\vec{\theta})\rangle$~\cite{per14}.
One example of such ansatz is the UCC-$2$, which uses the exponential of the second order expansion of the cluster operator:
\begin{align*}
|\psi(t_p^r,t_{pq}^{rs})\rangle=e^{T^{(2)}-T^{(2)\dag}}|\Phi_{\mathrm{ref}}\rangle,\\
T^{(2)}=\sum_{p,r}t_p^r\hat{f}^{\dag}_p\hat{f}_r+\sum_{p,q,r,s}t_{pq}^{rs}\hat{f}^{\dag}_p\hat{f}^{\dag}_q\hat{f}_r\hat{f}_s.
\end{align*}
After Trotterizing, this requires only weight-two or -four Majorana rotations to prepare.

\begin{figure}[tbh]
\includegraphics[width=\columnwidth]{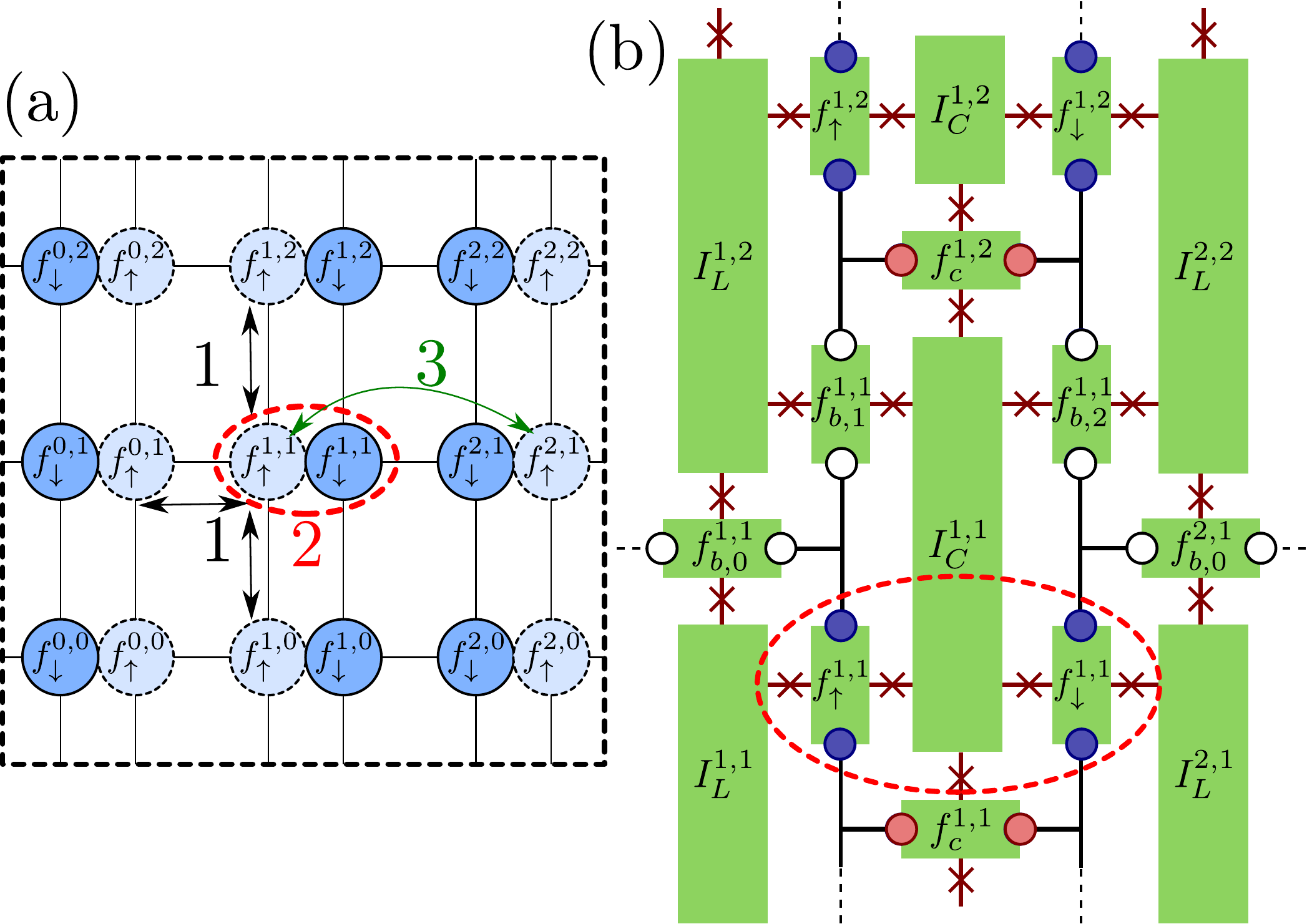}
\caption{\label{fig:2darchitecture} A 2d Majorana architecture to implement the Hubbard model on a square lattice. (a) A schematic description of the initial layout of the fermions (each of which is made of two Majoranas). Lines denote fermions separated by ancilla Majoranas only. Our scheme groups the $11$ Trotter steps into three stages as numbered, which are performed in series. (b) A physical architecture to support the schematic of (a). Wires on superconducting islands and T-junction symbols from Fig.~\ref{fig:4fold_rotation} have been removed to prevent cluttering; it is still assumed that all T-junctions are fully tunable. Majoranas are colored according to their designation; blue for system fermions, red for control ancillas, and white for braiding and phase ancillas. An example spin-$1/2$ fermion supported on four Majoranas (the minimum possible) is matched to (a)}
\end{figure}

When the Hamiltonian contains a small fraction of all possible second- or fourth-order terms, the lack of Jordan-Wigner strings gives our fermionic architecture an advantage over qubit-based implementations.
As an example, we consider the Hubbard model on a square lattice, with Hamiltonian
\begin{equation}
H=-t\sum_{\langle i,j\rangle,\sigma}\hat{f}^{\dag}_{i,\sigma}\hat{f}_{j \sigma} + U\sum_i\hat{n}_{i\uparrow}\hat{n}_{i\downarrow} - \mu\sum_{i\sigma}\hat{n}_{i\sigma}.
\end{equation}
Here $\sigma$ is a spin index, and the first sum is goes over the pairs of nearest neighbor lattice sites, while $t$, $\mu$, and $U$ are the model parameters~\cite{tasaki_hubbard_1998}. Rewriting the Hubbard model Hamiltonian in terms of Majorana operators $\hat{f}^{\dag}_{i\sigma}=\frac{1}{2}(\gamma_{\sigma,1}^i+i\gamma_{\sigma,2}^i)$ gives:
\begin{multline}
H=\frac{t}{2}\sum_{\langle i,j\rangle,\sigma}i\gamma_{\sigma,1}^{i}\gamma_{\sigma,2}^{j} + N(\frac{U}{4} -\mu)\\ + \frac{i}{4}(U-2\mu)\sum_{i,\sigma}\gamma_{\sigma,1}^i\gamma_{\sigma,2}^i-\frac{U}{4}\sum_i\gamma_{\uparrow,1}^i\gamma_{\uparrow,2}^i\gamma_{\downarrow,1}^i\gamma_{\downarrow,2}^i.\label{eq:Hubbard}
\end{multline}
This gives in total $11$ terms per site $i$ that need to be simulated for quantum phase estimation or unitary time evolution. In Fig.~\ref{fig:2darchitecture} we show a $2$d architecture that implements parallel application of Trotter steps across the entire lattice.
For unitary evolution, this scheme is $33\%$ dense, with $12$ Majoranas used per site with $2$ fermions. For parallel QPE we use an additional ancilla per site (following App.~\ref{app:QPE_ancilla}), making the scheme $50\%$ dense.
We detail the computation scheme for QPE in App.~\ref{app:HubbardCircuits}, achieving a $O(1)$ circuit depth per controlled unitary evolution step.
This should be compared first to the $O(N^{1/2})$ circuit depth in the case of a qubit implementation via a parallelized Jordan-Wigner transformation~\cite{Jiang17}.
This circuit depth can be reduced to $O(\log(N))$ if the Bravyi-Kitaev transformation~\cite{bravyi_fermionic_2002} is used instead, but at the cost of requiring dense qubit connectivity.
Separate encodings~\cite{Hav17,Whi16} also exist to reduce the circuit depth to $O(1)$, at a cost of doubling the required number of qubits.
It is likewise possible to achieve a similar $O(1)$ circuit depth, assuming the ability to couple a global resonator to every qubit in a superconducting architecture~\cite{Zhu17}.

\begin{figure*}[tbh]
\includegraphics[width=\textwidth]{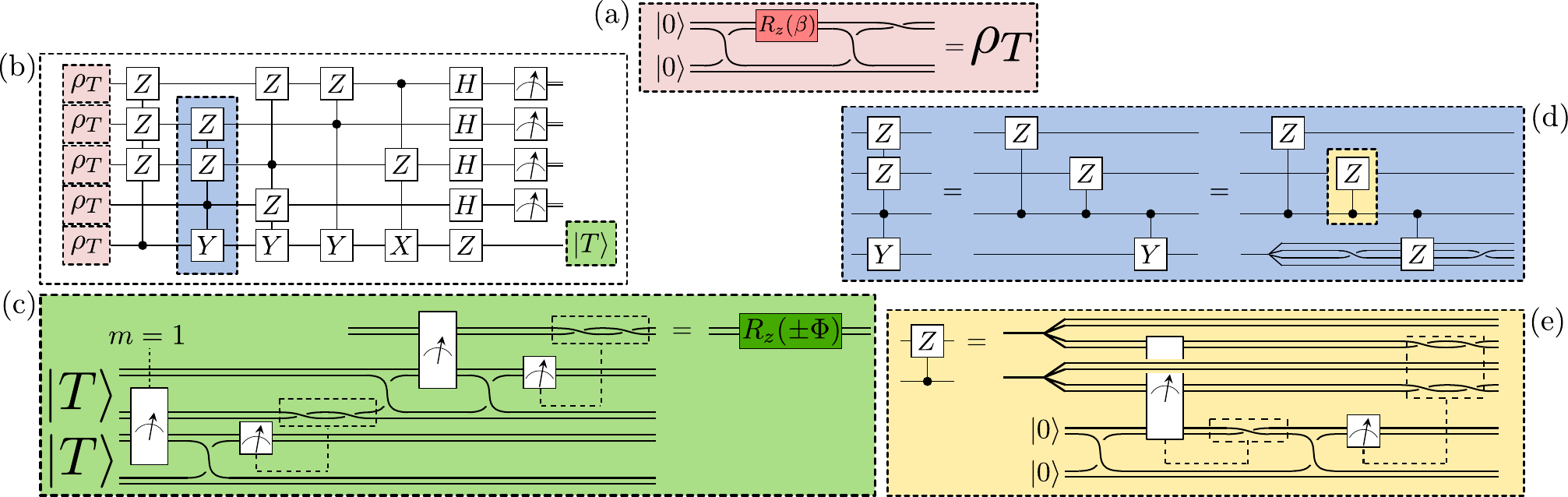}
\caption{\label{fig:MSD}Circuits for magic state distillation of a non-Clifford fermionic gate, following the scheme of~\cite{bra05}. (a) A noisy $\rho_T$ state is prepared with a single non-topologically protected gate. (b) 5 such-prepared states are distilled to give a single state with higher fidelity. (c) Two $|T\rangle$ states are consumed to perform a non-Clifford rotation of $\Phi=\frac{\pi}{12}$ on a single fermion, restoring universal quantum computation. This requires that the first measurement returns a value of $m=1$, otherwise a new pair of $|T\rangle$ states must be used. (d) To perform the state distillation protocol, we split the multi-qubit conditional gates into two-qubit controlled gates, and then into conditional-Z gates on the underlying fermions by braiding. (e) controlled Z gate: it may be performed by a circuit requiring braiding and correlated readout with a four-Majorana ancilla.}
\end{figure*}

The required ingredient for universal fermionic quantum computation---a Majorana rotation by an arbitrary angle $\theta$---is most simply implemented using an unprotected coupling between two Majoranas.
In a scalable architecture this gate needs to have increasingly higher fidelity so that it may be applied an arbitrary number of times without failure.
In Fig.~\ref{fig:MSD} we develop a high fidelity Majorana rotation using the magic state distillation protocol of~\cite{bra05} to perform fermionic gates.
In this procedure, we generate $5$ low-fidelity $|T\rangle=\cos(\beta)|0\rangle+e^{i\pi/4}\sin(\beta)|1\rangle$ states ($\cos(2\beta)=\frac{1}{\sqrt{3}}$) on four-Majorana qubits, then combine them to obtain a single higher fidelity $|T\rangle$ state on a qubit (assuming topologically-protected Clifford gates).
We then use an average of 3 distilled $|T\rangle$ states to perform a $\theta=\pm\frac{\pi}{12}$ Majorana rotation.
On average, this procedure requires $15$ noisy $|T\rangle$ states, 225 braidings and 66 measurements.
We furthermore use 20 Majoranas to make the 5 noisy $|T\rangle$ qubit states, due to the $|T\rangle$ state of a single fermion breaking parity conservation.

In summary, we have demonstrated a Majorana-based scheme for fermionic quantum computation.
We then adapted this scheme to simulate interacting fermionic Hamiltonians using both the QPE and VQC algorithms, and modified it to simulate the Hubbard model using a constant-depth circuit per time-evolution step.
While our fermionic scheme has advantages compared to using qubits, finding optimal circuit layouts for both a general purpose fermionic quantum computation and problem-specific ones, like the Hubbard model simulator remain an obvious point for further research.
Further, our implementation of magic state distillation is a direct translation of the original scheme, and it should be possible to find a smaller circuit operating only on fermions, for example using the minimal fermionic error correcting circuit of~\cite{Vijay17}.
A final open direction of further research is combining our circuits with quantum error correction~\cite{li_fault-tolerant_2017,Vijay17}, which would enable fault-tolerant fermionic quantum computation.

\acknowledgments
We have benefited from discussions with C.~W.~J.~Beenakker, F.~Hassler, B.~van~Heck, M.~Wimmer, M. Steudtner, and M. Munk-Nielsen. This research is supported by the Netherlands Organization for Scientific Research (NWO/OCW), as well as ERC Synergy and Starting Grants.

\bibliographystyle{apsrev4-1}
\bibliography{MajoranaQC}

\appendix
\section{Preparing extended ancilla qubits for quantum phase estimation}\label{app:QPE_ancilla}
\begin{figure}
\centering{
\includegraphics[width=0.75\columnwidth]{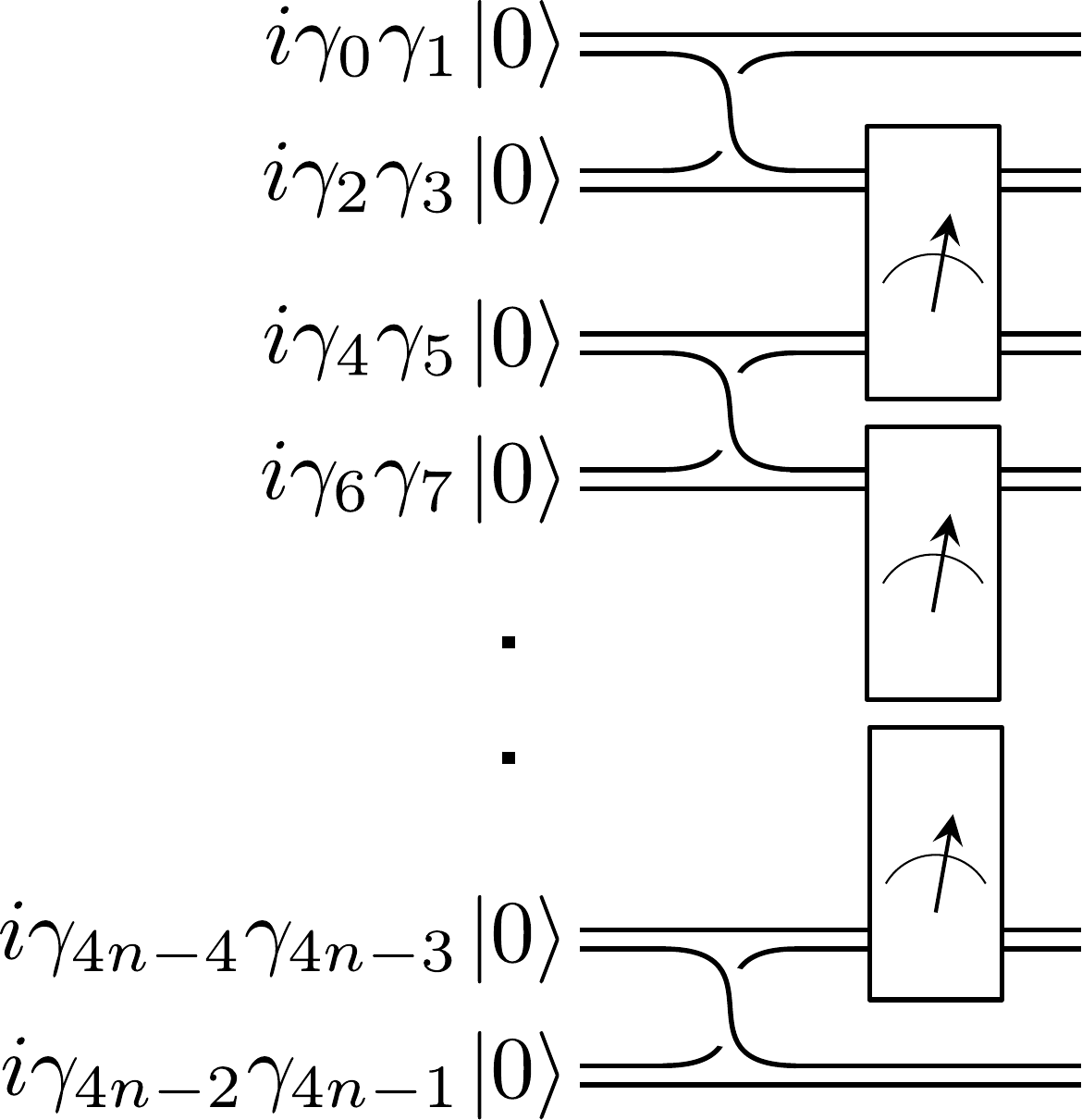}}
\caption{\label{fig:cat_state_prep}Circuit for preparing an extended cat state on a set of ancilla qubits with constant depth. The circuit need only be as local as the weight-four parity checks allow. Afterwards, any pair $\{\gamma_{2j},\gamma_{2j+1}\}$ of Majoranas may be used equivalently to perform a conditional Trotter step in QPE.}
\end{figure}

The QPE algorithm requires the application of a unitary operator conditional on an ancilla qubit, which naively would require each Trotter step to be performed in series as the ancilla qubit is passed through the system. The following method parallelizes the QPE algorithm at a cost of $O(N)$ ancilla qubits and a constant depth preparation circuit, which may well be preferable. We make this trade by preparing a large cat state on $4n$ Majoranas by the circuit in Fig.~\ref{fig:cat_state_prep}. First, we prepare $n\times 4$ Majoranas in the $\frac{1}{2}(|00\>+|11\>)$ state on Majoranas $\gamma_{4j}\gamma_{4j+1}\gamma_{4j+2}\gamma_{4j+3}$ for $j=0,\ldots,n-1$. Then, making the joint parity measurements $\gamma_{4j+2}\gamma_{4j+3}\gamma_{4j+4}\gamma_{4j+5}$ for $j=0,\ldots,n-2$ forces our system into an equal superposition of
\begin{equation}
\frac{1}{\sqrt{2}}\left(\left|\prod_{j=0}^{n-1}x_{2j}x_{2j+1}\right\>+\left|\prod_{j=0}^{n-1}\bar{x}_{2j}\bar{x}_{2j+1}\right\>\right),
\end{equation}
where $x_j\in\{0,1\}$ is the parity on the $j$th fermion ($\bar{x}=1-x$), and $x_{2j}\oplus x_{2j-1}$ is determined by the outcome of the joint parity measurement. This can then be converted to the GHZ state $\frac{1}{\sqrt{2}}(|00\ldots 0\>+|11\ldots 1\>)$ by braiding (or the value of $x_j$ can be stored and used to decide whether to rotate by $\theta$ or $-\theta$). The rotations to be performed for QPE may then be controlled by \emph{any} of the pairs of Majoranas defining a single fermion, and so we may spread this correlated ancilla over our system as required to perform rotations. As the interaction between ancilla qubits and system qubits is limited to a single joint parity measurement per Trotter step, we expect that although $n$ should scale as $O(N)$ to allow for parallelizing the circuit, the prefactor will be quite small.
At the end of the QPE circuit, we recover the required phase by rotating $\exp(i\frac{\pi}{4}\gamma_{4j+1}\gamma_{4j+2})$ for $j=0,\ldots,n-1$ and reading out the parity of all Fermions individually.
Starting from the state
\begin{equation}
\frac{1}{\sqrt{2}}\left( |00\ldots 0\>+e^{i\phi} |11\ldots 1\>\right),
\end{equation}
this prescription yields a $\cos^2(\phi/2)$ probability for the sum of all parities to be $0$ mod $4$.

\section{An algorithm to perform a Trotter step for a fully-connected fourth-order Hamiltonian in $O(N^3)$ time.}\label{app:TrotterN3}
We showed in the main text a compact circuit for a four-Majorana Trotter step that does not require Jordan-Wigner strings, and in App.~\ref{app:QPE_ancilla} we suggested a method to perform conditional evolution in parallel by using a large GHZ state for an ancilla qubit.
Assuming a Fermionic Hamiltonian on $N$ spin-orbitals with $4$th order terms, this would imply an $O(N^3)$ circuit depth for our QPE algorithm per Trotter step.
However, there is an additional complication; we need to ensure that we do not gain additional circuit depth from the requirement to bring sets of $4$ Majoranas close enough to perform this conditional evolution.
To show this, we consider a line of $N$ Majoranas $\gamma_1,\ldots\gamma_N$.
We allow ourselves at each timestep $t$ to swap a Majorana with its neighbour on the left or the right.
(Note that this is a simplification from our architecture where we may not directly swap initialized Majoranas, but this brings only an additional constant time cost.)
We wish to give an algorithm of length $O(N^3)$ such that for any set of four Majoranas $\{\gamma_i,\gamma_j,\gamma_k,\gamma_l\}$, there exists a timestep $t$ where these are placed consecutively along the line.
As demonstrated in~\cite{kivlichan_quantum_2017}, inverting the line by a bubblesort solves the equivalent problem for pairs $\{\gamma_i,\gamma_j\}$ in $O(N)$ time, and this may be quickly extended to the case of sets of four.
Let us consider the problem of forming all groups of $3$ Majoranas.
We divide our line into the sets $\Gamma_0=\{\gamma_i,i\leq N/2\}$, and $\Gamma_1=\{\gamma_i,i>N/2\}$. We then group neighboring pairs of elements in $\Gamma_1$ to form subsets, which we then pair with all elements in $\Gamma_0$ in $O(N)$ time by a reverse bubblesort.
Then, upon restoring to our previous position, we fix the position of elements of $\Gamma_0$, and perform a single iteration of the reverse bubblesort on the elements of $\Gamma_1$ to form new subsets of pairs.
Repeating this procedure until the second bubblesort has finished generates all subsets consisting of $2$ Majoranas in $\Gamma_1$ and $1$ from $\Gamma_0$ in $O(N^2)$ time.
All groups of $2$ Majoranas from $\Gamma_0$ and $1$ from $\Gamma_1$ may be given in the same manner.
Then, we may split the line in $2$, and reapply the above method on $\Gamma_0$ and $\Gamma_1$ separately to obtain all groups consisting of $3$ Majoranas within.
This final step takes $O((N/2)^2+(N/4)^2+(N/8)^2+\ldots)=O(N^2)$ time.
From here, it is clear how to proceed for groups of $4$.
We again divide our line into the sets $\Gamma_0$ and $\Gamma_1$, and split our problem into that of making all groups of $(m,4-m)$ Majoranas, where the first index denotes the number from $\Gamma_0$ and the second from $\Gamma_1$.
For $1\leq m\leq3$, we have an $O(N^{m-1})$ circuit to prepare all groups of $m$ Majoranas in $\Gamma_0$, a $O(N^{3-m})$ circuit to prepare groups of $4-m$ Majoranas in $\Gamma_1$, and an $O(N)$ bubblesort to pair all groups from $\Gamma_0$ and $\Gamma_1$.
These three steps must be looped within each other, giving a total time of $O(N^{m-1}N^{3-m}N)=O(N^3)$.
Finally, we perform the $m=0$ and $m=4$ case simultaneously by repeating this procedure on the sets $\Gamma_0$, which takes again $O(N^3)$ time by the arguments above.

\section{Details of parallel circuit for Hubbard model}\label{app:HubbardCircuits}
In this section we expand upon the proposal in Fig.~\ref{fig:2darchitecture} to perform QPE for the Hubbard model in constant time. This is a key feature of proposals for pre-error correcting quantum simulation~\cite{dal16}, and as such bears further detail.
There are $11$ terms in equation~\ref{eq:Hubbard} per site of our lattice, corresponding to $11$ Trotter steps that must be performed in series (as each circuit piece requires accessing a prepared ancilla and additional Majoranas for the controlled braiding). 
As part of these Trotter steps, we must move Majoranas to their appropriate islands for parity measurements, and leave sufficient space for the preparation of the controlled rotation gate.
We split the $11$ Trotter steps into $3$ stages, as indicated in Fig.~\ref{fig:2darchitecture}(a).
In the first stage, the Trotter steps corresponding to hopping terms between nearest neighbour fermions of the same spin are implemented, but only for those neighbours that are directly connected on the graph of Fig.~\ref{fig:2darchitecture}(a) (i.e.~those separated by a single braiding ancilla fermion).
In the second stage, the steps for onsite two and four fermion interactions are implemented.
From stage 2, as the qubits are being brought back to their resting position, the spin up and spin down fermions on each site have their locations exchanged.
This allows for the final two Trotter steps to be applied between fermions that are now locally connected, without the large overhead of bringing distant fermions together and then apart.
At the end of the unitary, the system is in a spin-rotated version of itself, and the order of Trotter steps for a second unitary evolution should be changed slightly to minimize braiding overhead.
In Table~\ref{tab:Hubbard_order}, we detail these three stages further.
In particular, we focus on the $10$ terms involving the fermion $f_\uparrow^{1,1}$, and the onsite interaction term for the fermion $f_{\downarrow}^{1,1}$.
For each term, we specify the location of all involved system Majoranas, parking spots for unused system Majoranas, the control ancilla, three braiding ancillas (for the implementation of the phase gate of Fig.~\ref{fig:4fold_rotation}), and which islands are involved in the parity measurement.
Each such set of operations should then be tessellated across the lattice by a translation of a unit cell and a spin rotation to generate $10$ parallelized Trotter steps for all fermions.
(For example, the hopping steps involving $f_\downarrow^{1,1}$ or $f_\uparrow^{1,2}$ are implemented in the operations from neighboring cells, and the hopping steps of $f_{\sigma}^{1,2}$ are reflected compared to those of $f_{\sigma}^{1,1}$, but those of $f_{\sigma}^{2,1}$ are not).
One should be careful then that this tessellation does not self-intersect, that all required qubits are connected to an island being measured, that the three braiding ancillas are connected in a way that allows for braiding, and that the measurement circuit does not isolate individual islands (which would cause them to dephase).
We assume that the conditional braidings on system Majoranas is performed as they move between configurations (or potentially cancelled), and so we do not account for these.
We also assume that our finite-sized lattice is surrounded by a common ground, and so parallel lines of coupled islands will maintain a common phase by connecting to this.
We have further found paths to hop Majoranas between their needed configurations and costed them in terms of the number of hoppings.
We make no claim that the found arrangement is optimal, and invite any interested readers to attempt to beat our score for an optimal braiding pattern.

\begin{table*}[p]
\begin{tabular}{c|c||c|c|c|c|c|c}\toprule
    \textbf{Stage}\;\;&\;\;\textbf{Hamiltonian}\;\; &\;\; \textbf{System} \;\; & \;\;\textbf{Parking}\;\; & \;\;\textbf{Control}\;\; & \;\;\textbf{Braiding}\;\; & \;\;\textbf{Measurement}\;\; & \;\; \textbf{Rearrangement} \;\;\\
    & \textbf{term} & \textbf{fermions} & \textbf{sites} & \textbf{ancilla} & \textbf{ancillas} & \textbf{island} & \textbf{cost}
    \\\hline
    \vspace{-0.2cm}& & & & & & & \\

    1 & $\frac{it}{2}\gamma_{\uparrow,1}^{1,1}\gamma_{\uparrow,2}^{0,1}$ & $f_{b,1}^{1,0}$ & $f_{b,0}^{1,0}$ & $f_{\uparrow}^{1,1}$ & $f_{\uparrow}^{0,1}$, $f_{c}^{0,1}$, $f_{b,2}^{0,0}$ & $I_L^{1,1} $ & (11) \\[5pt]\hline
    \vspace{-0.2cm}& & & & & & \\

    1 & $\frac{it}{2}\gamma_{\uparrow,1}^{0,1}\gamma_{\uparrow,2}^{1,1}$ & $f_{b,0}^{1,0}$ & $f_{b,1}^{1,0}$ & $f_{\uparrow}^{1,1}$ & $f_{\uparrow}^{0,1}$, $f_{c}^{0,1}$, $f_{b,2}^{0,0}$ & $I_L^{1,1} $ & 0 \\[5pt]\hline
    \vspace{-0.2cm}& & & & & & \\

    1 & $\frac{it}{2}\gamma_{\uparrow,1}^{1,1}\gamma_{\uparrow,2}^{1,2}$ & $f_\uparrow^{1,1}$ & $f_{b,1}^{1,1}$ & $f_c^{1,2}$ & $f_{b,2}^{1,1}$, $f_{b,0}^{2,1}$, $f_{\downarrow}^{1,1}$ & $I_C^{1,1}$ & 11+7 \\[5pt]\hline
    \vspace{-0.2cm}& & & & & & \\

    1 & $\frac{it}{2}\gamma_{\uparrow,1}^{1,2}\gamma_{\uparrow,2}^{1,1}$ & $f_{b,1}^{1,1}$ & $f_\uparrow^{1,1}$ & $f_c^{1,2}$ & $f_{b,2}^{1,1}$, $f_{b,0}^{2,1}$, $f_{\downarrow}^{1,1}$ & $I_C^{1,1}$ & 0 \\[5pt]\hline
    \vspace{-0.2cm}& & & & & & \\

    1 & $\frac{it}{2}\gamma_{\uparrow,1}^{1,1}\gamma_{\uparrow,2}^{1,0}$ & $f_{b,1}^{1,0}$ & $f_{\uparrow}^{1,0}$ & $f_c^{1,1}$ & $f_{b,2}^{1,0}$, $f_{b,0}^{2,0}$, $f_{\downarrow}^{1,0}$ & $I_C^{1,0}$ & 7+7 \\[5pt]\hline
    \vspace{-0.2cm}& & & & & & \\

    1 & $\frac{it}{2}\gamma_{\uparrow,1}^{1,0}\gamma_{\uparrow,2}^{1,1}$ & $f_{\uparrow}^{1,0}$ & $f_{b,1}^{1,0}$ & $f_c^{1,1}$ & $f_{b,2}^{1,0}$, $f_{b,0}^{2,0}$, $f_{\downarrow}^{1,0}$ & $I_C^{1,0}$ & 0\\[5pt]\hline
    \vspace{-0.2cm}& & & & & & \\

    2 & $\frac{i}{4} (U-2\mu)\gamma_{\uparrow,1}^{1,1}\gamma_{\uparrow,2}^{1,1}$ & $f_{c}^{1,1}$ & $f_{b,2}^{1,1}$ & $f_{\downarrow}^{1,1}$ & $f_{b,1}^{1,1}$, $f_{b,0}^{1,1}$, $f_{\uparrow}^{1,1}$ & $I_C^{1,1}$ & 7+6 \\[5pt]\hline
    \vspace{-0.2cm}& & & & & & \\

    2 & $\frac{i}{4} (U-2\mu)\gamma_{\downarrow,1}^{1,1}\gamma_{\downarrow,2}^{1,1}$ & $f_{b,2}^{1,1}$ & $f_{c}^{1,1}$ & $f_{\downarrow}^{1,1}$ & $f_{b,1}^{1,1}$, $f_{b,0}^{1,1}$, $f_{\uparrow}^{1,1}$ & $I_C^{1,1}$ & 0 \\[5pt]\hline
    \vspace{-0.2cm}& & & & & & \\

    2 & $-\frac{U}{4}\gamma_{\uparrow,1}^{1,1}\gamma_{\uparrow,2}^{1,1}\gamma_{\downarrow,1}^{1,1}\gamma_{\downarrow,2}^{1,1}$ & $f_{b,2}^{1,1}$, $f_{c}^{1,1}$ & & $f_{\downarrow}^{1,1}$ & $f_{b,1}^{1,1}$, $f_{b,0}^{1,1}$, $f_{\uparrow}^{1,1}$ & $I_C^{1,1}$ & 0 \\[5pt]\hline
    \vspace{-0.2cm}& & & & & & \\

    3 & $\frac{it}{2}\gamma_{\uparrow,1}^{1,1}\gamma_{\uparrow,2}^{2,1}$ & $f_{b,1}^{1,0}$ & $f_{b,0}^{1,0}$ & $f_{\uparrow}^{1,1}$ & $f_{\uparrow}^{0,1}$, $f_{c}^{0,1}$, $f_{b,2}^{0,0}$ & $I_L^{2,1} $ & 28+11 \\[5pt]\hline
    \vspace{-0.2cm}& & & & & & \\

    3 & $\frac{it}{2}\gamma_{\uparrow,1}^{2,1}\gamma_{\uparrow,2}^{1,1}$ & $f_{b,0}^{2,0}$ & $f_{b,2}^{1,0}$ & $f_{\downarrow}^{1,1}$ & $f_{\downarrow}^{2,1}$, $f_{c}^{2,1}$, $f_{b,1}^{2,0}$ & $I_L^{2,1} $ & 0 (+11) \\[5pt]\hline

\end{tabular}
\caption{\label{tab:Hubbard_order}Full scheme for an implementation of QPE on the Hubbard model (Eq.~\eqref{eq:Hubbard}), using the architecture in Fig.~\ref{fig:2darchitecture}. We specify a translatable layout for each Trotter step to be performed simultaneously, by specifying which sites should be used to store system fermions, control ancilla fermions, braiding ancilla fermions, and any additional fermions not used in this rotation (parking fermions). We further specify the island to be used for any joint parity readout. For each Trotter step we have costed the number of Majorana hoppings required to rearrange the system from its previous state. When these are written as a sum, the first term refers to restoring the configuration of Fig.~\ref{fig:2darchitecture} from the configuration required for the previous step, and the second to obtaining the configuration needed for the current step. Some steps require the same configuration as the previous step, and as such incur a $0$ rearrangement cost. The cost in brackets for the final step is the requirement to return the system to its shifted initial state (where up-spins and down-spins have been swapped). This may not be required, especially as the configuration for the final step and the initial steps are the same (modulo the swapping of the spins), and so repeated unitary evolution would not need this nor the rearrangement cost of the first step. This reduces the total rearrangement cost of the circuit to 85 Majorana hoppings.}
\end{table*}

\end{document}